\documentclass{elsart41}
\usepackage{graphics}
\usepackage{graphicx}
\usepackage{amssymb}
\begin{document}

\begin{frontmatter}



\title{Two-dimensional tetramer-cuprate Na$_{5}$RbCu$_{4}$(AsO$_{4}$)$_{4}$Cl$_{2}$: \ \ \  \ \ \ \ \ \ 
phase transitions and AFM order as seen by   $^{87}$Rb NMR}
%

\author[EE]{R. Stern\corauthref{Stern}},
\author[EE]{I. Heinmaa},
\author[EE]{A. Kriisa},
\author[EE]{E. Joon},
\author[EE]{S. Vija},
\author[OH]{J. Clayhold},
\author[SC]{M. Kartin-Ulutagay},
\author[SC]{X. Mo},
\author[SC]{W. Queen}, and
\author[SC]{S.-J Hwu}

\address[EE]{ NICPB, Akadeemia tee 23, EE12618 Tallinn, Estonia}
\address[OH]{Physics Department, Miami University, Oxford, OH 45056, USA}
\address[SC]{Department of Chemistry, Clemson University, Clemson, SC 29634, USA}
\corauth[Stern]{\textsl{Corresponding author stern@kbfi.ee,  tel: +372 513 22 88.}}

\begin{abstract}

We report  the $^{87}$Rb nuclear magnetic resonance (NMR) results  in a recently synthesized Na$_{5}$RbCu$_{4}$(AsO$_{4})$Cl$_{2}$. This complex novel two-dimensional (2D) cuprate is
an unique magnetic material, which contains layers of coupled
Cu$_{4}$O$_{4}$ tetramers.  
In zero applied magnetic field, it orders
antiferromagnetically via a second-order low-entropy phase
transition at $T_{N}$ = 15(1) K. 
We characterise the ordered state by $^{87}$Rb NMR, and
suggest for it a non-collinear rather than collinear arrangement
of spins. We discuss 
the properties of 
Rb nuclear site and point out the new structural phase transition(s)
around 74 K and 110 K.

\end{abstract}

\end{frontmatter}
The occurrence of high-temperature superconductivity in doped
spin-$1/2$ square planar antiferromagnets has stimulated the search
for new families of low-dimensional magnetic materials. We have
studied the compound Na$_{5}$RbCu$_{4}$(AsO$_{4})$Cl$_{2}$, whose
spin exchange interactions are confined to 2D layers.
The compound has a layered structure
comprised of square Cu$_{4}$O$_{4}$ tetramers \cite{Hwu}. The Cu ions are
divalent and the system behaves as a low-dimensional S = 1/2
antiferromagnet. Spin exchange in
Na$_{5}$RbCu$_{4}$(AsO$_{4})$Cl$_{2}$ appears to be
quasi-2D and nonfrustrated \cite{Clayhold}.

\begin{figure}[!ht]
\begin{center}
\includegraphics[width=0.50  \textwidth]{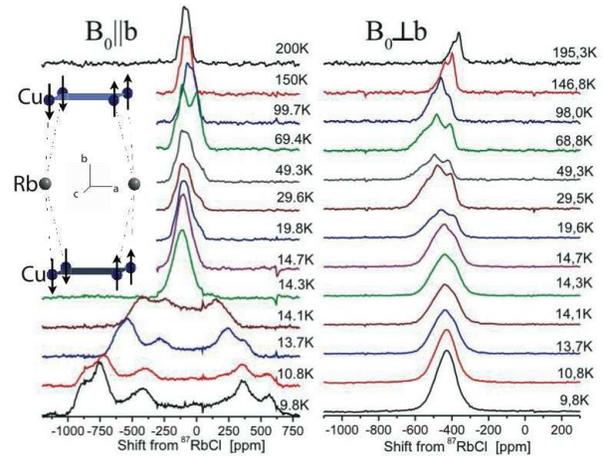}
\end{center}
\caption{The $T$-dependence of $^{87}$Rb central transition NMR spectra in B$_{0}$ = 14.1 T external field B$_{0} \parallel b$ (left) and (b) B$_{0} \bot b$ (right). Inset shows the position of Rb ions relative to Cu tetramers.} \label{fig1} 
\end{figure}

In this report we present $^{87}$Rb NMR results of the compound. The experiments are performed in B$_0$ =  8.45 T and 14.1T magnetic field  in a temperature ($T$)
range 4 K $<$T$<$300 K. A single crystal of Na$_{5}$RbCu$_{4}$(AsO$_{4})$Cl$_{2}$ (1 $\times$ 1 $\times$ 0.2 mm) was used.

\begin{figure}[!ht]
\begin{center}
\includegraphics[width=0.5\textwidth]{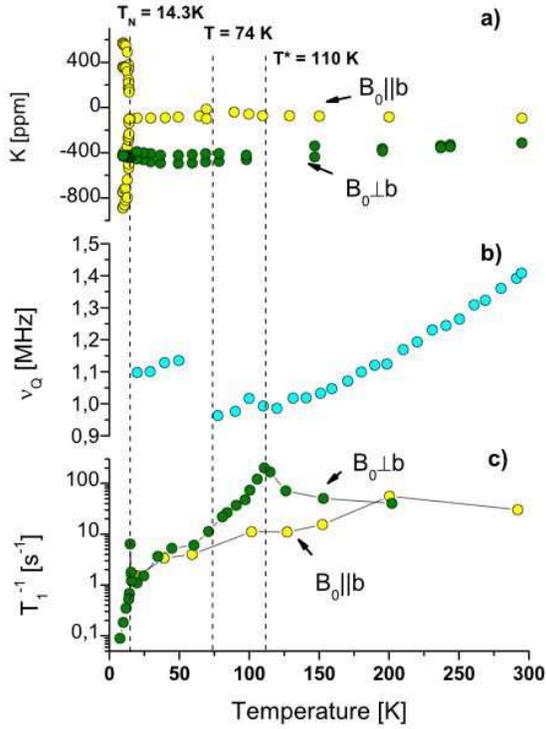}
\end{center}
\caption{The $T$-dependence of (a) the magnetic hyperfine shift K for B$_{0}$
$\parallel$ b-axis (B$_{0}$ = 14.1 T), (b)  the quadrupolar coupling parameter $\nu_Q = e^2 q Q/2h$, and (c) spin-lattice relaxation rate T$_{1}^{-1}$ (B$_{0}$= 8.45 T). Dotted lines note the $T_{N}$ and the high $T$ transitions at 74 K and 110 K.} \label{fig2}
\end{figure}
	 
The NMR spectra of $^{87}$Rb ($I = \frac{3}{2}$) central transition ($\pm \frac{1}{2} \Leftrightarrow \mp \frac{1}{2}$)  in Fig. \ref{fig1}(left) for the orientation of a crystal $B_{0}$
$\parallel b$ show a direct observation of the transition into the
AFM ordered phase below 14.3 K. 
The splitting of the resonance line corresponds to the local hyperfine field of $\pm$ 8 mT due to the static magnetic moments at the tetramers coppers. The spectra in Fig. \ref{fig1} (right) for $B_{0}$
$\perp b$ orientation do not show any measurable splitting in this $T$-range. From such orientation dependence one can conclude that in the AFM ordered phase the magnetic moments are aligned along the b-axis of the crystal, i.e. perpendicular to the tetramer plane. 
Rb sites in the lattice neighbor to four nearest coppers belonging to two sides of the two adjacent (along b) tetramers. These four coppers need to create a hyperfine field at Rb site in b-direction (or in opposite direction). A possible arrangement for that would be if the two magnetic moments of the tetramer sides along c axis have 
$\uparrow \uparrow$ or $\downarrow \downarrow$ alignment (insert of Fig. \ref{fig1}). That peculiar magnetic structure may result from strongly different superexchange couplings $J_a$ and $J_c$ between the coppers of a given tetramer \cite{Clayhold}. 

A closer look to the spectra in AFM phase indicates that the real magnetic structure of the compound is not simple at all. The complicated double-horn line shape with continuum of finite intensity between the peaks is a signature of an incommensurate structure. Generally, the NMR spectrum represents the distribution of magnetic hyperfine field at nuclei parallel to the external field. In a $\textit{non-collinear}$  incommensurate spin structure the hyperfine field should have a modulation, resulting in the continuum. The peaks in the spectra would then correspond to the extrema of that modulation. 

The separation between peaks in the spectrum in the ordered phase
grows steeply. We fitted peak shift $\nu$ to a power law
$\nu\approx $(T$_{N}$-T)$^{\beta}$ and obtained a critical
exponent $\beta$=0.25$\pm$ 0.05 and the transition temperature
T$_{N}$=14.8 K (14.3 K) for B$_{0}$=8.45 T (14.1 T). 

The $T$-dependence of the magnetic hyperfine shift K, the quadrupolar coupling value $\nu_Q = e^2 q Q/2h$, and the spin-lattice relaxation rate T$_{1}^{-1}$ are given  in Fig. \ref{fig2}.  Panel (a) shows a weak $T$-dependence of K above $T \geq T_N$ in both orientations. The splitting of the line in 
$B_{0} \bot b$ direcrion reflects a small anisotropy of K in $ac$-plane.
Below $T$ = 74 K the resonance lines in both directions show remarkable distortions which we attribute to the change of a local symmetry at Rb site. A clear indication of a symmetry change is the discontinuity of the $T$-dependence of the quadrupolar coupling constant around that $T$ ( Panel (b)). Since $\nu_Q$ describes local electric field gradient, it is highly sensitive to structural changes.
The $T$-dependence of spin-lattice relaxation rate T$_{1}^{-1}$
of $^{87}$Rb central transition shows also a N$\grave{e}$el temperature at
$T_{N}$=14.8 K and a broad peak around $T^*$=110K for B$_{0}$
$\bot$ b (Panel (c)). Similar $T$-dependence of T$_{1}^{-1}$ was measured in 14.1 T field. The relaxation is caused by local fluctuating magnetic fields that have to be aligned perpendicular
to the applied field to see the transitions. The same location of the maximum at 110 K in both fields suggests strongly that not the fluctuation spectrum but the crystal structure is changing at this $T$.
We do not know the resulting low-$T$ crystal structure in details yet, but the new structure may have crucial implications for the distinct magnetic order below $T_N$.

In conclusion, using $^{87}$Rb NMR we have shown the antiferromagnetic ordering and two high-$T$ phase transitions in Na$_{5}$RbCu$_{4}$(AsO$_{4})$Cl$_{2}$. The studies to clarify the nature of those phase transitions using NMR of $^{23}$Na,  $^{35}$Cl,  and $^{63,65}$Cu are currently in progress.

This work was supported by the Estonian Science Foundation and by the National Science Foundation (NSF) Grants DMR-0077321 and 0322905.

\end{document}